\documentclass[aps,twocolumn,showpacs,preprintnumbers,nofootinbib,prd,10pt,superscriptaddress]{revtex4-1}
\makeatletter
\def\l@subsubsection#1#2{}
\def\l@subsubsubsection#1#2{}
\makeatother

\usepackage{graphicx,amssymb,amsmath,amsthm,amsfonts,epsfig,epsf}
\usepackage{mathrsfs}
\usepackage{epstopdf}
\usepackage{mathtools}
\usepackage{bm}
\usepackage{enumerate}
\usepackage{tensor}
\usepackage{multirow}
\usepackage{url}
\makeatletter

\usepackage{ragged2e}

\usepackage[dvipsnames]{xcolor}
\usepackage[unicode]{hyperref}
\hypersetup{colorlinks=true, citecolor=MidnightBlue,
            linkcolor=MidnightBlue, urlcolor=MidnightBlue, linktocpage=true}
\usepackage{orcidlink}

\newcommand{\dd}{\mathrm{d}}

\usepackage{graphicx} 

\begin{document}
\title{Ruling out matter-induced bumps in black-hole effective potentials}
\author{Matteo Della Rocca}
\email{matteo.dellarocca@uniroma1.it}
\affiliation{Dipartimento di Fisica, Sapienza Universit\`a di Roma, Piazzale Aldo Moro 5, 00185, Roma, Italy}
\affiliation{INFN, Sezione di Roma, Piazzale Aldo Moro 2, 00185, Roma, Italy}

\author{Romeo Felice Rosato}
\email{romeofelice.rosato@uniroma1.it}
\affiliation{Dipartimento di Fisica, Sapienza Universit\`a di Roma, Piazzale Aldo Moro 5, 00185, Roma, Italy}
\affiliation{INFN, Sezione di Roma, Piazzale Aldo Moro 2, 00185, Roma, Italy}

\author{Paolo Pani}
\email{paolo.pani@uniroma1.it}
\affiliation{Dipartimento di Fisica, Sapienza Universit\`a di Roma, Piazzale Aldo Moro 5, 00185, Roma, Italy}
\affiliation{INFN, Sezione di Roma, Piazzale Aldo Moro 2, 00185, Roma, Italy}

\begin{abstract}
    Matter fields localized around black holes can produce small secondary bumps in the effective potentials governing black-hole perturbations, potentially leading to quasinormal-mode spectral instabilities and late-time echoes.
    Considering physically motivated models of anisotropic fluids surrounding a spherically symmetric black hole, 
    we show that the appearance of such a secondary bump is accompanied by either singular matter profiles or divergent characteristic speeds in the fluid. These features indicate that the corresponding matter configurations are unphysical.
    We further demonstrate that a secondary bump in the effective potential does not necessarily imply the existence of additional light rings in the spacetime. 
\end{abstract}

\maketitle
\section{Introduction}
The ringdown phase of compact-binary mergers provides a unique probe of the strong-field dynamics of gravity~\cite{Berti:2025hly}. As gravitational-wave detectors become increasingly sensitive and next-generation observatories come online~\cite{LIGOScientific:2025slb,ET:2019dnz,ET:2025xjr,LISA:2024hlh}, the quasinormal-mode~(QNM) spectrum of black holes~(BHs) is expected to become one of the primary tools for testing General Relativity and searching for new physics through BH spectroscopy~\cite{Berti:2025hly}.

Ringdown is traditionally modeled as a superposition of damped QNMs, whose complex frequencies are uniquely determined by the properties of the remnant BH. In practice, however, extracting these modes from GW signals is challenging, owing to ambiguities in the choice of the ringdown starting time and mode content~\cite{Capuano:2025kkl,Volkel:2025jdx}, possible nonlinear contributions~\cite{Cheung:2022rbm,Mitman:2022qdl}, and the risk of overfitting~\cite{Zhu:2023mzv,Baibhav:2023clw}. Beyond these practical challenges lies a more fundamental issue: due to the nonself-adjoint nature of the underlying boundary-value problem, the QNM spectrum is intrinsically spectrally unstable~\cite{Nollert:1996rf,Barausse:2014tra,Daghigh:2020jyk,Jaramillo:2020tuu}. Tiny perturbations of the background geometry, effective potential, or boundary conditions may induce large changes in the spectrum while leaving the corresponding spacetime and time-domain waveform almost unchanged. This phenomenon was first observed by Nollert~\cite{Nollert:1996rf}, who approximated the Regge--Wheeler potential by a sequence of small rectangular steps. Although the approximated potential accurately reproduced both the exact potential and the associated waveform, it yielded a dramatically different QNM spectrum.

The origin and physical implications of spectral instability have attracted considerable attention in recent years. Besides pseudospectral analyses of nonself-adjoint operators~\cite{Jaramillo:2020tuu,Destounis:2021lum}, a particularly influential approach has been to introduce localized bump-like deformations of the BH effective potential~\cite{Barausse:2014tra,Cheung:2021bol,Courty:2023rxk}. Such bumps provide a simple framework to mimic environmental matter distributions or near-horizon structures~\cite{Barausse:2014tra,Barausse:2014pra,Cardoso:2016rao,Maggio:2020jml,Cardoso:2019rvt,Maggio:2021ans,Cardoso:2024mrw,Courty:2023rxk}. Remarkably, these tiny deformations can completely reorganize the QNM spectrum while leaving the prompt ringdown signal in the time domain essentially unchanged~\cite{Berti:2022xfj,Rosato:2024arw,Yang:2024vor} and manifesting themselves only as subdominant late-time echoes~\cite{Barausse:2014tra,Cardoso:2016rao,Rosato:2024arw}. Estimates of the magnitude of matter-induced corrections to BH effective potentials, and of the environmental conditions under which they may become relevant, were investigated in Ref.~\cite{Boyanov:2024fgc}.

At the same time, other scattering observables, including greybody factors and cross sections, remain comparatively stable under the same deformations~\cite{Rosato:2024arw,Oshita:2024fzf,Torres:2023nqg}. These results naturally raise a fundamental question: \emph{Can the bump-like potentials commonly employed to study spectral instability actually arise from physically admissible matter configurations?}

In this work we address this question by considering a general static, spherically symmetric anisotropic fluid surrounding a Schwarzschild BH. Working perturbatively in the matter distribution, we determine the stress-energy tensor required to generate an arbitrarily prescribed localized deformation of the Regge--Wheeler potential and investigate whether the corresponding matter configuration satisfies basic physical requirements such as regularity and causality. 

We first show that a localized bump does not necessarily imply the existence of additional light rings: matter-induced bumps that exist for perturbations at finite multipole number may disappear in the eikonal limit, corresponding to a spacetime with a single unstable photon orbit, as for a standard Schwarzschild BH in vacuum. 

We then investigate the commonly adopted phenomenological assumption that the effect of the surrounding matter can be captured solely by an additive deformation of the Regge--Wheeler potential, leaving the remaining structure of the perturbation equation unchanged. For a generic anisotropic fluid, this requirement does not uniquely determine the matter configuration, and a suitable choice of the residual freedom can formally reproduce a prescribed localized bump. However, when this assumption is applied to physically motivated fluid models, such as strictly static, irrotational, and isotropic fluids, it leads to strong constraints that preclude a regular and causal realization of the desired deformation. In particular, localized bumps either cannot be supported by a nontrivial regular matter configuration or require divergent characteristic speeds.
Finally, we further show that the same issue arises for smooth regularizations of infinitesimally thin matter shells, which have also been employed as models of spectral instability~\cite{Leung:1997was,Leung:1999iq,Barausse:2014tra,Laeuger:2025zgb}. 

Our results do not call into question spectral instability itself, which is an intrinsic mathematical property of BH perturbations. Rather, they show that one of the most widely used models for investigating this phenomenon does not, in general, admit a physically viable realization in terms of surrounding matter. More broadly, our analysis highlights the limitations of modeling physically motivated environmental effects through a deformation of the effective potential alone. A self-consistent treatment must also incorporate the matter-induced changes to the background geometry, the perturbation operator, the matter source terms, and possibly the associated boundary conditions.

All our main results are derived analytically and were independently confirmed through numerical calculations. We use $G=c=1$ units.
%
\section{Two-parameter expansion of the spacetime}
%
We consider the linear response to an external perturbation of a static, spherically symmetric dirty BH whose geometry is continuously connected to the Schwarzschild solution. Accordingly, we write the background metric as
\begin{equation}\label{eq:metric_generic}
g^{(0)}_{\mu\nu}
=
g^{\rm Sch}_{\mu\nu}
+\epsilon\,\tilde g_{\mu\nu}\,,
\end{equation}
where $\epsilon$ parametrizes the departure from Schwarzschild and $\tilde g_{\mu\nu}$ describes the matter-induced deformation of the geometry; the Schwarzschild limit is recovered as $\epsilon\to0$. By construction, $\tilde g_{\mu\nu}$ preserves staticity and spherical symmetry. Through Einstein's equations, this deformation is sourced by a nonvanishing stress-energy tensor $T_{\mu\nu}$ describing the matter surrounding the BH.
We then consider linear perturbations of this background, introducing a second expansion parameter $\zeta$ according to
\begin{equation}\label{eq:metric_generic_pert}
g_{\mu\nu}=g^{(0)}_{\mu\nu}+\zeta \ h_{\mu\nu}\ ,
\end{equation}
where $\zeta\ll1$ is a bookkeeping parameter. 
Thus, the spacetime is described by two independent expansion parameters: $\epsilon$ characterizes the deformation of the background from Schwarzschild, while $\zeta$ controls the amplitude of the perturbation.
Owing to the spherical symmetry of $g^{(0)}_{\mu\nu}$, $h_{\mu\nu}$ can be decomposed into spherical harmonics and separated into axial (odd-parity) and polar (even-parity) sectors~\cite{Ferrari:2020nzo}. 
We focus on the axial sector for simplicity, although we expect similar results for the polar sector.
%
\subsection{Linearized Tolman-Oppenheimer-Volkoff equations}
For the background, we consider the general spherically symmetric spacetime,
\begin{equation}\label{eq:metric}
    \dd s^2=-A(r)\dd t^2+\frac{1}{B(r)} \dd r^2+r^2(\dd\theta^2+\sin^2\theta \ \dd\phi^2) \ .
\end{equation}
To describe the effect of a small overdensity (like a small bump, or a tiny matter shell) around a Schwarzschild BH with mass $M_{\rm BH}$, we assume $A=f+\epsilon\delta A$ and $B=f+\epsilon\delta B$, with  $f=1-2M_{\rm BH}/r$. The functions $\delta A$ and $\delta B$ account for small corrections due to a nonvanishing stress-energy tensor,
\begin{equation}
    T_{\mu}^\nu={\rm diag}(-\rho(r),p_r (r),p_t(r),p_t(r)) \ ,
\end{equation}
where $\rho=\epsilon\delta\rho$ is the density, $p_r=\epsilon\delta p_r$ is the radial pressure, and $p_t=\epsilon\delta p_t$ is the tangential pressure. 
At zeroth order in $\epsilon$ the field equations are trivially satisfied.
At linear order in $\epsilon$, Einstein's equations reduce to 
\begin{subequations}\label{eq:einsteins}
\begin{align}
\delta B'&=-8 \pi r \delta \rho -\frac{\delta B(r)}{r}\ ,\label{eq:dB}\\
\delta A'&=-\frac{r \delta B-2 M_{\rm BH} \delta A}{r^2 f}+8\pi  r \delta p_r\ , \label{eq:dA}\\
\delta p_r'&=-\frac{M_{\rm BH} \delta \rho}{r^2f}+\frac{(3 M_{\rm BH}-2 r) \delta p_r}{r^2 f}+\frac{2 \delta p_t}{r} \label{eq:dpt}\ ,
\end{align}
\end{subequations}
where a prime denotes differentiation with respect to $r$. 
To close this system for the five variables $(\delta A, \delta B, \delta\rho, \delta p_r,\delta p_t)$, two additional relations are required. These are typically provided by equations of state relating the tangential and radial pressures to the energy density. In the following we will assume that such relations are barotropic, i.e. $\delta p_r=\delta p_r(\delta \rho)$ and $\delta p_t=\delta p_t(\delta \rho)$.
Within this local barotropic description, the derivatives
$c_{s,r}^2=\dd\delta p_r/\dd\delta\rho$ and
$c_{s,t}^2=\dd\delta p_t/\dd\delta\rho$ define the corresponding squared characteristic speeds. Causality and stability against radial and tangential gradient instabilities therefore require
$0\leq c_{s,i}^2\leq1$, with $i=r,t$.
%
\subsection{A perturbative approach to the Regge--Wheeler equation}
%
As shown in Appendix~A of Ref.~\cite{Zhao:2026eti}, in the axial sector, each multipole $(\ell,m)$ is governed by a single master equation, which generalizes the Regge--Wheeler equation~\cite{Regge:1957td} to the dirty-BH background (cf.~Eq.~A32 in Ref.~\cite{Zhao:2026eti}):
\begin{equation}\label{eq:RW}
    (\hat{\mathcal O}-V^\ell)\psi^{\ell m}=S_{\rm fluid}^{\ell m}\ , \quad \hat{\mathcal{O}}=\partial_{r_\star}^2+\omega^2 \ ,
\end{equation}
where $r_\star$ is the tortoise coordinate associated with the background spacetime, $V^\ell$ and $\psi^{\ell m}$ denote the modified Regge--Wheeler potential and master function, respectively, while
$S_{\rm fluid}^{\ell m}$, defined in Eq.~A34 in Ref.~\cite{Zhao:2026eti}, reads 
\begin{equation}\label{eq:S_def}
    S^{\ell m}_{\rm fluid}=-16 \pi (p_r-p_t) \sqrt{A^3 B}\frac{X^{\ell m}}{r}
\end{equation}
in our conventions.
Here, $X^{\ell m}$ characterizes the axial component of the matter perturbation, obtained by projecting the perturbation of the vector $k^\mu$, i.e., the spatial vector orthogonal to the fluid four-velocity, onto the vector harmonics on the two-sphere.
In principle, $X^{\ell m}$ is an undetermined function whose form depends on the assumptions made about the nature of the fluid. 
In general, it has the form
\begin{equation}\label{eq:X}
    X^{\ell m}=f^{-1/2}F^{\ell m}(\psi,\psi',\dots,\omega)\ ,
\end{equation}
where $F$ is a generic linear function of the Regge--Wheeler  variable, the matter perturbation functions, and their spatial derivatives (see Appendix~A in Ref.~\cite{Zhao:2026eti} for details).

The above treatment is nonperturbative in $\epsilon$.
In the regime $\epsilon\ll 1$, the background spacetime differs perturbatively from the Schwarzschild geometry, and the Regge--Wheeler operator, master function, and potential can be expanded to linear order in $\epsilon$ as~\cite{DOnofrio:2026ulh,Zhao:2026eti}
\begin{subequations}\label{eq:RW_exp}
\begin{align}
    \hat{\mathcal O}&=\hat{\mathcal O}_{(0)}+\epsilon\, \hat{\mathcal O}_{(1)} \ ,\\
    \psi^{\ell m}&=\psi^{\ell m}_{(0)}+\epsilon \,\psi^{\ell m}_{(1)} \ , \\
    V^\ell&=V^\ell_{(0)}+\epsilon \,\delta V^\ell \ , 
\end{align}
\end{subequations}
where the subscript $(0)$ refers to quantities in vacuum. The function $S^{\ell m}_{\rm fluid}$, instead, is already ${\cal O}(\epsilon)$ (see Eq.~\eqref{eq:S_def}). 
Solving Eq.~\eqref{eq:RW} order by order in $\epsilon$, we get two coupled differential equations 
\begin{align}
   &\left(\hat{\mathcal O}_{(0)}-V^\ell_{(0)}\right)\psi^{\ell m}_{\rm(0)}=0 \ , \label{eq:RW_vacuum}\\ 
    &\left(\hat{\mathcal{O}}_{(0)}-V_{(0)}^\ell\right)\psi^{\ell m}_{(1)}=-\hat{\mathcal O}_{(1)}\psi^{\ell m}_{(0)}+\delta V^\ell\psi^{\ell m}_{(0)}+S_{\rm fluid}^{\ell m} \ .\label{eq:first_order}
\end{align}
Equation~\eqref{eq:RW_vacuum} is the standard Regge--Wheeler equation in vacuum~\cite{Regge:1957td}, where $V^\ell_{(0)}$ has a single peak corresponding to the unstable light ring in the eikonal limit~\cite{Ferrari:2020nzo} and whose amplitude grows with $\ell$, while Eq.~\eqref{eq:first_order} accounts for (perturbative) deviations from the vacuum case. We note that the vacuum Regge--Wheeler function $\psi^{\ell m}_{(0)}$ acts as a source for the function $\psi^{\ell m}_{(1)}$.
Using the line element in Eq.~\eqref{eq:metric}, together with the vacuum Regge--Wheeler equation~\eqref{eq:RW_vacuum} and Eqs.~\eqref{eq:S_def} and~\eqref{eq:X}, we can obtain the explicit expressions for the terms appearing on the right-hand side of Eq.~\eqref{eq:first_order}:
\begin{align}\label{eq:01}
\begin{split}
  &\hat{\mathcal{O}}_{(1)}\psi_{(0)}=-(\delta A+\delta B)f^{-1}(\omega^2-V_{(0)})\psi_{(0)}\\
  &\quad +\left[\frac{1}{2} f \left(\delta A'+\delta B'\right)-\frac{M_{\rm BH}}{r^2}\left( \delta A+\delta B\right)\right]\partial_r\psi_{(0)} \ ,
\end{split}\\ 
\begin{split}
&\delta V^\ell=\frac{3 f \delta B}{r^2}+\frac{\delta A (r \ell  (\ell +1)-6 M_{\rm BH})}{r^3}\\
&\qquad-4 \pi f (\delta p_r-\delta \rho +4
   \delta \sigma)\ ,
\end{split}\\
&S_{\rm fluid}^{\ell m}=-\frac{16\pi  f^{3/2} \delta \sigma }{r}F^{\ell m}(\psi_{(0)},\psi_{(0)}',\dots,\omega) \ ,\label{eq:generic_S}
\end{align}
where we introduced $\delta\sigma=\delta p_r-\delta p_t$.
In what follows, we shall
omit the harmonic indices $(\ell,\ m)$ to avoid clutter.

\section{Bumpy potentials and matter profiles}\label{sec:int_scheme}
%
%
Inspired by previous studies~\cite{Cheung:2021bol,Courty:2023rxk}, we investigate whether there exist matter profiles whose sole effect on the 
Regge--Wheeler equation is to generate a localized bump in the effective potential. This requirement directly leads to a set of constraints that effectively act as equations of state and close the system in Eqs.~\eqref{eq:einsteins}.

%
%
\subsection{Digression: matter-induced bumps and light rings}\label{sec:lightring}
%
Before presenting our main argument, it is useful to clarify the relation between secondary bumps in the Regge--Wheeler potential and additional light rings. Reference~\cite{Cardoso:2024mrw} argued that producing a secondary bump requires a nonperturbative deformation of the Schwarzschild geometry, because such a feature is associated with the appearance of an additional stable light ring. The key point, however, is that this correspondence applies to features that persist in the eikonal limit. A secondary maximum at fixed, finite multipole number $\ell$ need not have a light-ring counterpart.

For the metric in Eq.~\eqref{eq:metric}, a circular null orbit is located at $r=\bar r$, where~\cite{Chandrasekhar:1985kt,Ferrari:1984zz}
\begin{equation}
\label{eq:LR_condition}
    \bar r A'(\bar r)=2A(\bar r)\ .
\end{equation}
Expanding this condition to linear order in $\epsilon$ gives
\begin{equation}
\label{eq:LR_exp}
    \frac{2\left(3M_{\rm BH}-\bar r\right)}{\bar r^2}
    +\epsilon\left[
        \delta A'(\bar r)
        -\frac{2\delta A(\bar r)}{\bar r}
    \right]
    =0\ .
\end{equation}
At zeroth order, Eq.~\eqref{eq:LR_exp} admits the Schwarzschild light ring
$\bar r_0=3M_{\rm BH}$. Assuming that $\delta A$ and its relevant derivatives
remain regular and perturbative, we write
$\bar r=\bar r_0+\epsilon\,\delta\bar r$. Substituting this expansion into
Eq.~\eqref{eq:LR_exp}, we obtain
\begin{equation}
\label{eq:LR_shift}
    \delta\bar r
    =
    \frac{\bar r_0^2}{2}\delta A'(\bar r_0)
    -\bar r_0\delta A(\bar r_0)\ .
\end{equation}
Thus, within a regular perturbative expansion, the Schwarzschild light ring
is continuously displaced rather than split into multiple light rings.
Equation~\eqref{eq:LR_shift} is a local statement about the continuation of
the nondegenerate Schwarzschild root. Additional roots could only arise in
regions where the perturbative correction, or its derivatives, is no longer
uniformly small; such configurations lie outside the perturbative regime
considered here. In this sense, the result is consistent with
Ref.~\cite{Cardoso:2024mrw}, which identifies the formation of additional
light rings as a genuinely nonperturbative effect.

However, this conclusion does not exclude the appearance of a secondary maximum in
the Regge--Wheeler potential at finite $\ell$. The correspondence between
extrema of the perturbation potential and circular null orbits is an
eikonal statement: at large $\ell$, the leading part of the
Regge--Wheeler potential is controlled by the null-geodesic potential, and
its extrema approach the light-ring locations. The height of the primary
peak therefore grows as $\ell(\ell+1)$. A matter-induced feature that grows
more slowly with $\ell$, or that is present only for a restricted set of
multipoles, becomes negligible relative to the primary peak as
$\ell\rightarrow\infty$. Such a feature may generate a secondary maximum
for finite $\ell$ while disappearing from the eikonal potential. The
light-ring correspondence is then recovered asymptotically without the
formation of an additional circular null orbit.

The argument of Ref.~\cite{Cardoso:2024mrw} applies when the secondary
feature remains finite relative to the primary peak in the eikonal limit.
Under that assumption, a persistent secondary extremum is indeed tied to
an additional light ring and therefore requires a nonperturbative
deformation of the geometry. The matter-induced bumps considered here
belong to a different regime: they are perturbative, finite-$\ell$
features that need not survive as $\ell\rightarrow\infty$.

We therefore conclude that a secondary bump in the Regge--Wheeler
potential is not, by itself, evidence for an additional stable light ring.
This observation motivates the analysis below, in which we investigate
whether physically admissible matter distributions can generate such
finite-$\ell$ bumps within a perturbative deformation of the Schwarzschild
geometry.
%

\subsection{Matter-induced bumps for generic fluids}
\label{sec:match}
%
We consider matter configurations whose only effect on the axial perturbation equation is to produce an additive deformation of the Regge--Wheeler potential. This restriction does not capture the most general matter-induced modification of the perturbation dynamics, which may also alter the differential operator or introduce an independent source.
Nevertheless, it reproduces the phenomenological setup commonly adopted in studies of localized deformations of BH potentials~\cite{Cheung:2021bol,Courty:2023rxk}.

Within our perturbative framework, implementing this setup requires Eq.~\eqref{eq:first_order} to take the form
\begin{equation}
\label{eq:RW_with_bump}
    \left(
        \hat{\mathcal O}_{(0)}-V^\ell_{(0)}
    \right)\psi_{(1)}
    =
    \delta V_{\rm bump}\,\psi_{(0)}\ ,
\end{equation}
where $\delta V_{\rm bump}$ depends on the radial coordinate but is
independent of the mode frequency. Comparing Eq.~\eqref{eq:RW_with_bump}
with Eq.~\eqref{eq:first_order}, we see that the combination
\begin{equation}
    -\hat{\mathcal O}_{(1)}\psi_{(0)}+S_{\rm fluid}
\end{equation}
must be proportional to $\psi_{(0)}$ and must contain neither derivatives of $\psi_{(0)}$ nor terms proportional to $\omega^2\psi_{(0)}$.
As shown by Eqs.~\eqref{eq:01}--\eqref{eq:generic_S}, this requirement strongly constrains the admissible axial matter perturbation $F$ introduced
in Eq.~\eqref{eq:X}.

Restricting attention to terms that are linear in $\psi_{(0)}$ and its first radial derivative, the required cancellations can be implemented with the ansatz
\begin{equation}
\label{eq:F}
    F
    =
    \left(x\omega^2+z\right)\psi_{(0)}
    +y\psi'_{(0)}\ ,
\end{equation}
where $x=x(r)$, $y=y(r)$, and $z=z(r)$. The functions $x$ and $y$ control, respectively, the contributions proportional to $\omega^2\psi_{(0)}$ and $\psi'_{(0)}$, whereas $z$ contributes directly to the effective potential deformation. Note that higher-order derivatives are not considered as they can be eliminated by using the zeroth-order field equations.

Substituting Eq.~\eqref{eq:F} into Eq.~\eqref{eq:first_order} and requiring
the coefficients of $\omega^2\psi_{(0)}$ and $\psi'_{(0)}$ to vanish gives two algebraic conditions. For nonvanishing anisotropic stress $\delta\sigma\neq0$, these conditions uniquely determine $x$ and $y$:
\begin{subequations}
\label{eq:xy}
\begin{align}
    x &=
    \frac{r\left(\delta A+\delta B\right)}
         {16\pi f^{5/2}\delta\sigma}\ ,
    \\
    y &=
    \frac{
        2M_{\rm BH}\left(\delta A+\delta B\right)
        -r^2f\left(\delta A'+\delta B'\right)
    }
    {32\pi r f^{3/2}\delta\sigma}\ .
\end{align}
\end{subequations}
The function $z(r)$ remains unconstrained by these cancellations. The special case $\delta\sigma=0$ cannot be inferred by taking the limit of Eqs.~\eqref{eq:xy} and must instead be considered directly at the level of Eq.~\eqref{eq:first_order}, as we shall do later on.

With the choices in Eqs.~\eqref{eq:xy}, the first-order perturbation
equation assumes the desired form~\eqref{eq:RW_with_bump}, with
\begin{equation}
\label{eq:Vbump_generic_fluid}
\begin{split}
    \delta V_{\rm bump}
    ={}&
    -\frac{16\pi f^{3/2}\delta\sigma\,z}{r}
    -4\pi f
        \left(\delta p_r-\delta\rho+4\delta\sigma\right)
    \\
    &-
    \frac{\left[\ell(\ell+1)-3\right]\delta B}{r^2}\ .
\end{split}
\end{equation}
Equation~\eqref{eq:Vbump_generic_fluid} is therefore not an independently
prescribed potential: once $\delta V_{\rm bump}$ and $z$ are specified, it
acts as an additional constitutive constraint relating the matter variables
to the metric perturbations.

Together, Eqs.~\eqref{eq:Vbump_generic_fluid} and
\eqref{eq:einsteins} provide four relations for the five background perturbations
$\delta\rho$, $\delta p_r$, $\delta p_t$, $\delta A$, and $\delta B$.
The function $z(r)$ also remains free. Consequently, the generic-fluid system is underdetermined, and this residual freedom may formally be used to engineer matter profiles that reproduce a prescribed bump.
We do not expect these formally constructed matter profiles to correspond to physically relevant matter models. In Sec.~\ref{sec:relevant_fluids}, we instead consider fluid models of astrophysical relevance and show that, within the perturbative framework adopted here, no admissible matter profile can modify the Regge--Wheeler equation solely through the addition of a bump to the effective potential.
%
\subsection{Bumpy potentials for physically relevant fluids}\label{sec:relevant_fluids}
%
For physically motivated matter models, including irrotational and strictly static fluids, the matter term $S_{\rm fluid}$ contains neither terms proportional to $\omega^2\psi_{(0)}$ nor terms proportional to $\psi'_{(0)}$~\cite{Zhao:2026eti}. Therefore, any terms proportional to $\omega^2\psi_{(0)}$ or $\psi'_{(0)}$ appearing on the right-hand side of Eq.~\eqref{eq:first_order} must originate entirely from $-\hat{\mathcal O}_{(1)}\psi_{(0)}$.
From Eq.~\eqref{eq:01}, both contributions vanish if and only if
\begin{equation}
\label{eq:eos1}
    \delta A=-\delta B\ .
\end{equation}
This condition is purely geometrical and does not depend on the detailed form of the stress-energy tensor. To linear order in $\epsilon$, it is equivalent to requiring that the tortoise coordinate retain its vacuum Schwarzschild form.

Substituting Eq.~\eqref{eq:eos1} into Eq.~\eqref{eq:dA} and combining the
result with Eq.~\eqref{eq:dB}, we obtain the additional algebraic constraint
\begin{equation}
\label{eq:algebraic3}
    \delta p_r
    =
    \delta\rho+\frac{\delta B}{4\pi r^2 f}\ .
\end{equation}
Equations~\eqref{eq:eos1} and~\eqref{eq:algebraic3} considerably restrict the matter configurations capable of producing a pure deformation of the Regge--Wheeler potential. In the following subsections, we examine these constraints for several astrophysically motivated matter models. The case
of an anisotropic fluid with vanishing radial pressure is treated separately in Sec.~\ref{sec:vanishing_pressure}, where it arises as a special case of
a more general result.

\subsubsection{Strictly static fluid}\label{sec:static}
%
For a strictly static fluid, the axial matter perturbation vanishes identically, so that $S_{\rm fluid}=0$ and hence $z=0$. Substituting
Eqs.~\eqref{eq:eos1} and~\eqref{eq:algebraic3} into
Eq.~\eqref{eq:first_order} and comparing the result with
Eq.~\eqref{eq:RW_with_bump}, we obtain
\begin{equation}
\label{eq:eos2}
    \delta V_{\rm bump}
    =
    16\pi f\left(\delta p_t-\delta\rho\right)
    -
    \frac{\ell(\ell+1)+2}{r^2}\delta B\ .
\end{equation}
Equations~\eqref{eq:eos1},~\eqref{eq:algebraic3}, and~\eqref{eq:eos2}, together with Eqs.~\eqref{eq:dpt} and~\eqref{eq:dB}, form a closed system for the five functions $\delta A$, $\delta B$, $\delta p_t$, $\delta p_r$, and $\delta\rho$.
The first three equations are algebraic, while the last two are ordinary differential equations. Therefore, once a profile $\delta V_{\rm bump}$ is prescribed, the corresponding matter and metric perturbations are determined after suitable boundary conditions are imposed.

Eliminating $\delta A$, $\delta p_r$, and $\delta p_t$ from this system, we obtain an evolution equation for the density perturbation:
\begin{equation}
\label{eq:drho}
    \delta\rho'
    =
    \frac{\delta V_{\rm bump}}{8\pi r f}
    +
    \frac{
        r f\left[\ell(\ell+1)+4\right]+2M_{\rm BH}
    }
    {8\pi r^4 f^2}\,\delta B
    +
    \frac{f+1}{r f}\,\delta\rho\ .
\end{equation}
Together with Eq.~\eqref{eq:dB}, this gives a coupled system of
first-order differential equations for $\delta\rho$ and $\delta B$.

It is convenient to express this system in terms of the perturbed mass
function $\delta m$, defined by
\begin{equation}
\label{eq:delta_m_definition}
    \delta m=-\frac{r\,\delta B}{2}\ .
\end{equation}
Equation~\eqref{eq:dB} then becomes
\begin{equation}
\label{eq:delta_m_density}
    \delta m'=4\pi r^2\delta\rho\ .
\end{equation}
Using Eqs.~\eqref{eq:drho} and~\eqref{eq:delta_m_density}, we find the
following second-order equation for $\delta m$:
\begin{equation}
\label{eq:mass_wave}
\begin{split}
    \delta m''
    ={}&
    \frac{3f+1}{r f}\,\delta m'
    -
    \frac{
        2M_{\rm BH}
        +r f\left[\ell(\ell+1)+4\right]
    }
    {r^3f^2}\,\delta m
    \\
    &+
    \frac{r\,\delta V_{\rm bump}}{2f}\ .
\end{split}
\end{equation}
For a prescribed potential deformation, this equation can be integrated subject to two boundary conditions.

To determine the admissible boundary conditions, we examine Eq.~\eqref{eq:mass_wave} near spatial infinity and the event horizon. Assuming that $\delta V_{\rm bump}$ decays sufficiently rapidly in both limits, the leading homogeneous solutions behave as
\begin{equation}
\label{eq:mass_asymptotics}
    \delta m \sim
    \begin{cases}
        r^{5/2}
        \left(
            c_1 r^{-i\beta}
            +c_2 r^{i\beta}
        \right),
        & r\rightarrow\infty,
        \\[1ex]
        r f
        \left[
            \hat c_1
            +\hat c_2
            \ln\left(\dfrac{r}{2M_{\rm BH}}-1\right)
        \right],
        & r\rightarrow 2M_{\rm BH},
    \end{cases}
\end{equation}
where $\beta=\frac{1}{2} \sqrt{4\ell(\ell+1)-9}$.
For gravitational perturbations, $\ell\geq2$, and both independent
asymptotic modes grow as $r^{5/2}$. An asymptotically admissible solution
therefore requires
\begin{equation}
    c_1=c_2=0 \ ,
\end{equation}
or, equivalently,
\begin{equation}
\label{eq:infinity_boundary_conditions}
    \delta m\rightarrow0 \ ,
    \qquad
    \delta m'\rightarrow0
    \qquad
    \text{as}\quad r\rightarrow\infty\  .
\end{equation}
These conditions ensure that the matter perturbation does not modify the asymptotic mass and are consistent with the requirement that the tortoise coordinate be the same as in Schwarzschild.

Near the horizon, regularity instead requires
\begin{equation}
    \hat c_2=0 \ ,
\end{equation}
since the logarithmic branch makes
$\delta\rho=\delta m'/(4\pi r^2)$ divergent at
$r=2M_{\rm BH}$. The asymptotic and horizon requirements thus impose three
independent conditions on the second-order inhomogeneous
Eq.~\eqref{eq:mass_wave}. For a generic prescribed
$\delta V_{\rm bump}$, they cannot all be satisfied simultaneously.

Indeed, integrating inward from spatial infinity with
Eq.~\eqref{eq:infinity_boundary_conditions} fixes the solution completely
but generically excites the logarithmic mode at the horizon, producing a
divergent density perturbation. Conversely, selecting the horizon-regular
solution and integrating outward generically excites the growing modes in
Eq.~\eqref{eq:mass_asymptotics}, violating asymptotic admissibility. A
globally regular solution could exist only for specially tuned potential
profiles satisfying an additional compatibility condition.

We therefore conclude that a generic, localized deformation of the
Regge--Wheeler potential cannot be generated as the sole perturbative
effect of a strictly static anisotropic fluid while maintaining both
horizon regularity and asymptotic admissibility. This obstruction does not
depend on the existence of a secondary maximum: it applies to any
prescribed deformation $\delta V_{\rm bump}$ that decays sufficiently
rapidly at both boundaries.

\subsubsection{Irrotational fluid}
%
For an irrotational fluid, the axial matter source takes the
form~\cite{Zhao:2026eti}
\begin{equation}
\label{eq:S_irrotational}
    S_{\rm fluid}
    =
    16\pi f\,\delta\sigma\,\psi_{(0)}\ .
\end{equation}
Comparison with Eqs.~\eqref{eq:generic_S} and~\eqref{eq:F} shows that this
corresponds to
\begin{equation}
    x=y=0 \ ,
    \qquad
    z=-r f^{-1/2}\ .
\end{equation}
As in the static case, we require
Eq.~\eqref{eq:first_order}, supplemented by
Eqs.~\eqref{eq:S_irrotational},~\eqref{eq:eos1}, and
\eqref{eq:algebraic3}, to reduce to the pure-potential form
\eqref{eq:RW_with_bump}. This requirement gives
\begin{equation}
\label{eq:dB_sol_analytical}
    \delta B
    =
    -\frac{r^2}{(\ell-1)(\ell+2)}
    \,\delta V_{\rm bump}\ .
\end{equation}
Thus, a prescribed potential deformation fixes the radial metric
perturbation algebraically. Equation~\eqref{eq:eos1} then determines
$\delta A$, while Eqs.~\eqref{eq:algebraic3},~\eqref{eq:dpt}, and
\eqref{eq:dB} determine the matter variables. Explicitly,
\begin{subequations}
\label{eq:matter_analytical_solution}
\begin{align}
    \delta\rho
    &=
    \frac{
        r\,\delta V_{\rm bump}'
        +3\delta V_{\rm bump}
    }
    {8\pi(\ell-1)(\ell+2)}\ ,
    \\
    \delta p_r
    &=
    \frac{
        r^2f\,\delta V_{\rm bump}'
        +(r-6M_{\rm BH})\delta V_{\rm bump}
    }
    {8\pi(\ell-1)(\ell+2)r f}\ ,
    \label{eq:dpr_sol_an}
    \\
    \delta p_t
    &=
    \left(
        1+\frac{2M_{\rm BH}^2}{r^2f^2}
    \right)
    \frac{\delta V_{\rm bump}}
         {8\pi(\ell-1)(\ell+2)}
    \nonumber\\
    &\quad+
    \frac{
        (2r-5M_{\rm BH})\delta V_{\rm bump}'
    }
    {8\pi(\ell-1)(\ell+2)f}
    +
    \frac{
        r^2\delta V_{\rm bump}''
    }
    {16\pi(\ell-1)(\ell+2)}\ .
    \label{eq:dpt_sol_an}
\end{align}
\end{subequations}
These expressions formally associate an irrotational-fluid configuration
with any prescribed profile $\delta V_{\rm bump}$. We now show, however,
that a localized profile satisfying the required asymptotic conditions
generically produces a singular effective radial sound speed.

Using Eq.~\eqref{eq:dB_sol_analytical} and the definition
$\delta m=-r\delta B/2$, we obtain
\begin{equation}
    \delta m
    =
    \frac{r^3}{2(\ell-1)(\ell+2)}
    \,\delta V_{\rm bump}\ .
\end{equation}
Requiring the tortoise coordinate to be the same as in Schwarzschild implies that the matter distribution does not change the asymptotic mass, $\delta m\rightarrow0$ as $r\rightarrow\infty$, and therefore that $\delta V_{\rm bump}$ decays sufficiently fast,
\begin{equation}
\label{eq:Vbumpinfinity}
    \delta V_{\rm bump}=o(r^{-3})\ ,
    \qquad
    r\rightarrow\infty\ .
\end{equation}
For a locally barotropic radial response, the effective radial sound speed along the matter profile is
\begin{equation}
\label{eq:csrirrotational}
    c_{s,r}^2
    =
    1+
    \frac{
        4M_{\rm BH}r\,\delta V_{\rm bump}
        -2fr^3\delta V_{\rm bump}'
    }
    {
        f^2
        \dfrac{\dd}{\dd r}
        \left(r^4\delta V_{\rm bump}'\right)
    }\ .
\end{equation}
We now show that this quantity necessarily becomes singular outside the
outermost extremum of a localized bump.

Let $r_0$ denote the outermost extremum of
$\delta V_{\rm bump}$, and suppose first that it is a positive maximum.
Because the potential approaches zero and has no further extrema for
$r>r_0$, we have
\begin{equation}
    \delta V_{\rm bump}>0 \ ,
    \qquad
    \delta V_{\rm bump}'<0 \ ,
    \qquad
    r>r_0\ .
\end{equation}
There must then exist at least one radius $r_1>r_0$ such that
\begin{equation}
\label{eq:vanishing_density_gradient}
    \left.
    \frac{\dd}{\dd r}
    \left(r^4\delta V_{\rm bump}'\right)
    \right|_{r=r_1}
    =0\ .
\end{equation}
To see this, define $h(r)=r^4\delta V_{\rm bump}'(r)$. At the maximum, $h(r_0)=0$, whereas $h(r)<0$ for $r>r_0$. Suppose, by contradiction, that $h'$ never vanishes in this region. By continuity, $h'$ must then be strictly negative throughout $r>r_0$, and hence $h$ is strictly decreasing. Therefore, for any fixed $R>r_0$,
\begin{equation}
h(r)\leq h(R)<0 \ ,
\qquad r\geq R\ .
\end{equation}
It follows that
\begin{equation}
\delta V_{\rm bump}'(r)
\leq \frac{h(R)}{r^4} \ .
\end{equation}
Using $\delta V_{\rm bump}(r)\to0$ as $r\to\infty$ and integrating from $r$ to infinity, we obtain
\begin{equation}\label{eq:cond_irr}
\delta V_{\rm bump}(r)
=
-\int_r^\infty \delta V_{\rm bump}'(x)\,dx
\geq -\frac{h(R)}{3r^3} \ .
\end{equation}
Since $h(R)<0$, Eq.~\eqref{eq:cond_irr} is incompatible with the asymptotic behavior in Eq.~\eqref{eq:Vbumpinfinity}. Therefore, $h'$ must vanish at least once at some finite radius $r_1>r_0$, which proves Eq.~\eqref{eq:vanishing_density_gradient}.

Thus, the denominator of the second term in
Eq.~\eqref{eq:csrirrotational} vanishes at $r=r_1$. Its numerator, by contrast, is
strictly positive:
\begin{equation}
    4M_{\rm BH}r\,\delta V_{\rm bump}
    -2fr^3\delta V_{\rm bump}'
    >0 \ ,
    \qquad
    r>r_0\ .
\end{equation}
Therefore, $c_{s,r}^2$ diverges at $r=r_1$. 

The same reasoning applies
when the outermost extremum is a negative minimum. In that case,
$\delta V_{\rm bump}<0$ and $\delta V_{\rm bump}'>0$ beyond the
extremum, so the numerator is strictly negative while the denominator
again vanishes at some finite radius.

We conclude that any sufficiently smooth, localized potential deformation with a nonzero extremum leads to a divergence in the effective radial sound speed of the associated irrotational-fluid configuration. Such a divergence renders the configuration incompatible with both causality and the absence of radial gradient instabilities. Hence, an irrotational fluid cannot provide a physically admissible realization of the bump-like potential deformations considered here.

\begin{figure}
    \centering
\includegraphics[width=\linewidth]{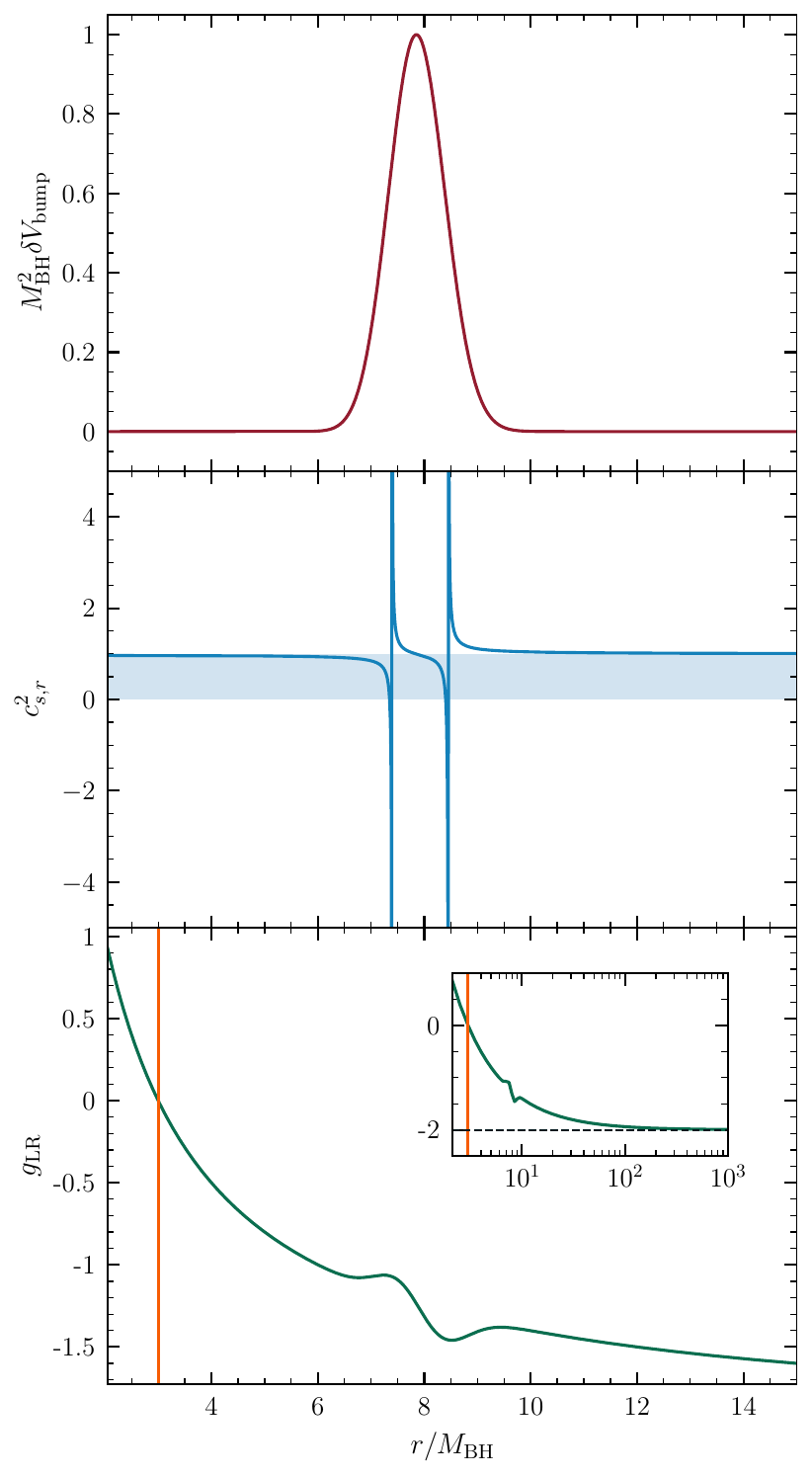}
        \caption{\small{
        Representative irrotational-fluid configuration with $\epsilon=10^{-3}$ and $\ell=2$. The upper panel shows the bump profile defined in Eq.~\eqref{eq:V_bump_example}, which has a maximum at $r\simeq 7.85 M_{\rm BH}$. The middle panel shows the squared radial sound speed from Eq.~\eqref{eq:csrirrotational}, which diverges at $r\simeq7.40M_{\rm BH}$ and  $r\simeq8.45M_{\rm BH}$. The shaded band marks the physically admissible interval $0\leq c_{s,r}^2\leq1$, corresponding to stability against radial gradient perturbations and subluminal propagation. The lower panel shows the light-ring function $g_{\rm LR}$ defined in Eq.~\eqref{eq:LR_condition_irrotational}; its zeros determine the light-ring radii. For the parameters considered here, the only zero occurs at $r\simeq3M_{\rm BH}$, as indicated by the vertical orange line. The inset displays the same function over a wider radial range, and the horizontal dashed line marks its asymptotic value, $g_{\rm LR}\rightarrow-2$.
        }}
    \label{fig:example}
\end{figure}

As an illustration, the upper panel of Fig.~\ref{fig:example} shows a
Gaussian deformation localized in the tortoise coordinate,
\begin{equation}
\label{eq:V_bump_example}
    \delta V_{\rm bump}
    =
    M_{\rm BH}^{-2}
    \exp\left[
        -M_{\rm BH}^{-2}
        \left(r_\star-10M_{\rm BH}\right)^2
    \right]\ ,
\end{equation}
whose center, $r_\star=10M_{\rm BH}$, corresponds to
$r\simeq7.85M_{\rm BH}$. We set $\ell=2$. The middle panel displays the
effective radial sound speed obtained from
Eq.~\eqref{eq:csrirrotational}, which diverges at
$r\simeq7.40M_{\rm BH}$ and $r\simeq8.45M_{\rm BH}$. The blue shaded
region identifies the interval in which
\begin{equation}
    0\leq c_{s,r}^2\leq1 \ ,
\end{equation}
so that the radial response is both free of gradient instabilities and
subluminal.

For completeness, the lower panel of Fig.~\ref{fig:example} shows the
corresponding light-ring function
\begin{equation}
\label{eq:LR_condition_irrotational}
    g_{\rm LR}
    =
    \frac{
        \epsilon r^3\delta V_{\rm bump}'
    }
    {(\ell+2)(\ell-1)}
    +
    2\left(
        \frac{3M_{\rm BH}}{r}-1
    \right)\ ,
\end{equation}
whose zeros determine the light-ring locations. This expression follows
from Eqs.~\eqref{eq:LR_exp},~\eqref{eq:eos1}, and
\eqref{eq:dB_sol_analytical}. For the illustrative choice
$\epsilon=10^{-3}$, $g_{\rm LR}$ has only one zero, located at
$\bar r\simeq3M_{\rm BH}$; the displacement from the Schwarzschild value
is negligible at the numerical precision shown. The inset displays
$g_{\rm LR}$ over a wider radial interval, while the horizontal dashed
line marks its asymptotic value, $g_{\rm LR}\rightarrow-2$.

This example illustrates both conclusions derived above: an irrotational fluid cannot support a bump deformation of the Regge--Wheeler potential without developing a divergent effective sound speed, while the bump considered for a finite multipole $\ell$ does not generate an additional light ring.
%
\subsubsection{Isotropic fluid}\label{sec:isotropic}
%
For an isotropic fluid, $\delta\sigma=0$, and hence $S_{\rm fluid}$ vanishes independently of the form of $F$. The bump condition is therefore identical to that obtained for a strictly static fluid, given in Eq.~\eqref{eq:eos2}. Imposing the equation of state $\delta p_r=\delta p_t=\delta p$ and using Eq.~\eqref{eq:algebraic3}, we obtain the same analytical solution for $\delta B$ as in the irrotational case, namely Eq.~\eqref{eq:dB_sol_analytical}.

The corresponding expressions for $\delta\rho$, $\delta p_r$, and $\delta p_t$ are consequently those given in Eq.~\eqref{eq:matter_analytical_solution}. Isotropy imposes the additional constraint $\delta p_r=\delta p_t$, so Eqs.~\eqref{eq:dpr_sol_an} and~\eqref{eq:dpt_sol_an} must coincide. This requirement yields a second-order ordinary differential equation for $\delta V_{\rm bump}$, whose general solution is
\begin{equation}\label{eq:deltaV_sol}
\begin{split}
\delta V_{\rm bump}&=
\frac{c_1 f}{M_{\rm BH} r}
\left(1-\frac{3M_{\rm BH}}{r}\right)
\\
&+\frac{c_2 f}{M_{\rm BH}^2}
\left(
1-\frac{3M_{\rm BH}}{r}
-\frac{27M_{\rm BH}^2}{r^2}
\right)
\\
&+\frac{8c_2 f}{M_{\rm BH}r}
\left(1-\frac{3M_{\rm BH}}{r}\right)
\ln\left(\frac{r}{2M_{\rm BH}}-1\right)\,,
\end{split}
\end{equation}
where $c_1$ and $c_2$ are dimensionless integration constants.

The asymptotic behavior of this solution is incompatible with that of a localized bump. In particular,
\begin{equation}
\delta V_{\rm bump}
=
\frac{c_2}{M_{\rm BH}^2}
+\mathcal{O}\left(\frac{\ln r}{r}\right)\,,
\qquad r\to\infty\,.
\end{equation}
Thus, if $c_2\neq0$, the potential approaches a nonzero constant. Setting $c_2=0$ removes this constant contribution, but the remaining solution decays only as
\begin{equation}
\delta V_{\rm bump}
\sim
\frac{c_1}{M_{\rm BH}r}\,,
\qquad r\to\infty\,.
\end{equation}
Therefore, the only solution with the required localized asymptotic behavior is the trivial one, $c_1=c_2=0$. We conclude that an isotropic fluid cannot generate a nontrivial localized bump while leaving the remaining structure of the Regge--Wheeler equation unchanged.
\section{Density extrema and causality}\label{sec:matter_distribution}
So far, our discussion of localized deformations of the effective potential relied on imposing specific restrictions on the tortoise coordinate. It is therefore natural to ask whether the conclusions reached above persist once these assumptions are relaxed.

Before addressing this question, we establish a simple and fairly general result that will prove useful in the remainder of the paper. Within the linearized framework adopted here, neither a fluid with vanishing radial pressure nor an isotropic fluid can support a smooth density profile with a nonvanishing extremum without developing a divergent characteristic speed.

To see this, let $\delta\rho(r)$ be an arbitrary smooth density perturbation. For a fluid with vanishing radial pressure,
\begin{equation}
\delta p_r=0\ ,
\end{equation}
the linearized Einstein equations~\eqref{eq:einsteins} imply
\begin{equation}
\delta p_t=
\frac{M_{\rm BH}}{2rf}\,\delta\rho \, ,
\label{eq:pt_pr0}
\end{equation}
whereas the radial sound speed vanishes identically. For a barotropic fluid, the tangential sound speed is instead given by
\begin{equation}
c_{s,t}^2=
\frac{\delta p_t'}{\delta\rho'} \, ,
\end{equation}
which, upon using Eq.~\eqref{eq:pt_pr0}, becomes
\begin{equation}
c_{s,t}^2=
\frac{M_{\rm BH}}{2rf}
-
\frac{M_{\rm BH}}{2r^2f^2}
\frac{\delta\rho}{\delta\rho'} \, .
\label{eq:cst_pr0_general}
\end{equation}
Suppose now that the density profile possesses a nonvanishing extremum at $r=r_0$. Since $\delta\rho'(r_0)=0$ while $\delta\rho(r_0)\neq0$, Eq.~\eqref{eq:cst_pr0_general} shows that $c_{s,t}^2$ necessarily diverges as $r\to r_0$. Therefore, any smooth density profile with a nonzero extremum is incompatible with a regular causal fluid description when the radial pressure vanishes. This result is consistent with the nonperturbative analysis of Ref.~\cite{Datta:2023zmd}, where similar pathologies were found for several density profiles surrounding a BH.

The same conclusion also holds for isotropic fluids, for which $\delta p_t=\delta p_r=\delta p$. As shown in Appendix~\ref{app:isotropic}, any nonvanishing density extremum likewise leads to a divergent sound speed, again signaling the breakdown of a causal fluid description.
%
\subsection{Bump matter profile with vanishing radial pressure}\label{sec:vanishing_pressure}
An interesting application of the previous general result is to rule out bumps in the effective potential supplied by fluids with vanishing radial pressure. 
For a fluid with vanishing radial pressure, i.e.,
$\delta p_r=0$, Eqs.~\eqref{eq:pt_pr0}
and~\eqref{eq:Vbump_generic_fluid} fix the density
$\delta\rho$ as 
\begin{equation}\label{eq:deltarho_ani}
    \delta \rho
    =
    \frac{
    \left(\ell^2+\ell-3\right)\delta B
    +r^2\delta V_{\rm bump}
    }{
    4\pi\left(2M_{\rm BH}f^{1/2}z+r^2\right)
    } \, .
\end{equation}
The system in Eq.~\eqref{eq:einsteins} then reduces to a
pair of equations for $\delta A$ and $\delta B$, namely
\begin{subequations}\label{eq:eqs_ani}
\begin{align}
    \delta A'
    &=
    \frac{2M_{\rm BH}\delta A-r\delta B}{r^2f}
    \,,
    \label{eq:eqs_ani_deltaA}\\
    \delta B'
    &=
    -\frac{
    2r\left[
    \left(\ell^2+\ell-3\right)\delta B
    +r^2\delta V_{\rm bump}
    \right]
    }{
    2M_{\rm BH}f^{1/2}z+r^2
    }
    -\frac{\delta B}{r}
    \, .
    \label{eq:dB_ani}
\end{align}
\end{subequations}
We now show that this system cannot support a localized
bump without violating causality.

First, we note that, if $\delta V_{\rm bump}$ is localized
at $r\gg2M_{\rm BH}$ and the term proportional to
$\delta B$ in Eq.~\eqref{eq:deltarho_ani} is subleading (i.e., for finite values of $\ell$),
it induces a corresponding localized feature in the density,
\begin{equation}
    \delta \rho
    \to
    \frac{
    r^2\delta V_{\rm bump}
    }{
    4\pi\left(2M_{\rm BH}f^{1/2}z+r^2\right)
    } \, .
\end{equation}
Thus, provided that the denominator is regular and varies
slowly across the localization region, the density inherits
the bump independently of the detailed form of $z$, although
its amplitude generally depends on $z$.

\paragraph{Case $z=0$.}
A fully analytical result can be obtained for $z=0$.
In this case, the general solution of
Eq.~\eqref{eq:dB_ani} is~\cite{Morse:1946met}
\begin{align}
\delta B
\notag&=
r^{-2\ell(\ell+1)+5}
\big[
c_1
\\&-2\int_{2M_{\rm BH}}^r
\dd x\,
x^{2(\ell+2)(\ell-1)}
\delta V_{\rm bump}(x)
\big] \, .
\end{align}
The integration constant $c_1$ is fixed by the behavior
of the metric perturbation at the horizon. Indeed, the
homogeneous contribution gives a finite, nonzero value of
$\delta B$ at $r=2M_{\rm BH}$. Substitution into Eq.~\eqref{eq:eqs_ani_deltaA} then produces a logarithmic
divergence in $\delta A'$ unless $c_1=0$.

Using $c_1=0$ in Eq.~\eqref{eq:deltarho_ani}, we obtain
\begin{equation}\label{eq:delta_rho_sol}
\begin{split}
\delta& \rho
={}
\frac{1}{4\pi}\delta V_{\rm bump}
\\
&-
\frac{\ell^2+\ell-3}{2\pi r^{2\ell(\ell+1)-3}}
\int_{2M_{\rm BH}}^r
\dd x\,
x^{2(\ell-1)(\ell+2)}
\delta V_{\rm bump}(x)
\, .
\end{split}
\end{equation}
Let us first assume that $\delta V_{\rm bump}$ is a positive localized bump. On its near-horizon tail, $\delta V_{\rm bump}$ increases monotonically from zero. Therefore, for $2M_{\rm BH}\leq x\leq r$, one has $\delta V_{\rm bump}(x)\leq\delta V_{\rm bump}(r)$, so that the integral in Eq.~\eqref{eq:delta_rho_sol} is at most of order $(r-2M_{\rm BH})\,\delta V_{\rm bump}(r)$. Since the remaining radial prefactor is finite at the horizon, the integral contribution is suppressed by an additional factor of $r-2M_{\rm BH}$ with respect to the first term, and is therefore subleading as $r\to2M_{\rm BH}^{+}$. 
At spatial infinity, we can define 
\begin{equation}
    I=\int_{2M_{\rm BH}}^r\dd x\,x^{2(\ell-1)(\ell+2)}\delta V_{\rm bump}(x)>0\,.
\end{equation}
Asymptotically, as long as $\delta V_{\rm bump}$ decays faster than $1/r^3$, the term involving $I$ determines the asymptotic sign of $\delta \rho$ in Eq.~\eqref{eq:delta_rho_sol}. Indeed, if $\delta V_{\rm bump}\sim A r^{-p}$, the integral contribution either decays more slowly than $\delta V_{\rm bump}$ or has the same $r^{-p}$ scaling with a larger coefficient in magnitude.

Thus, we obtain the following asymptotic behaviors
\begin{equation}
   \delta \rho \sim  \begin{cases}
      0^+ \quad\text{for}\quad r\to2M_{\rm BH}^+  \,,\\
      0^-  \quad\text{for}\quad r\to \infty\,.
    \end{cases}
\end{equation}

Hence, any nontrivial positive localized bump necessarily
produces a density profile that changes sign. Since
$\delta\rho$ is continuous and approaches zero at both
boundaries with opposite signs, it must possess at least
one nonvanishing extremum at a finite radius
$r_0\in(2M_{\rm BH},\infty)$. The tangential sound speed
therefore diverges at $r_0$.

For a negative localized bump, the signs of the two
asymptotic limits are reversed, and the same conclusion
holds.

\paragraph{Generic regular $z$.}
The previous result can be generalized to a nonvanishing function
$z(r)$. We assume that $z$ is regular throughout the BH
exterior and remains finite at the horizon. Regularity of the density profile implies
\begin{equation}
D_z= r^2+2M_{\rm BH}f^{1/2}z\neq0
\end{equation}
for all $r>2M_{\rm BH}$ as a zero of $D_z$ would generically produce a singularity in Eq.~\eqref{eq:deltarho_ani}. Since $D_z\to4M_{\rm BH}^2>0$ at the horizon, continuity then implies that $D_z>0$ throughout the exterior. 

Defining $a_\ell=\ell^2+\ell-3$, Eq.~\eqref{eq:dB_ani} can be written as
\begin{equation}
\delta B'+\left(\frac{1}{r}+\frac{2a_\ell r}{D_z}\right)\delta B=-\frac{2r^3}{D_z}\delta V_{\rm bump}\,.
\end{equation}
Introducing
\begin{equation}
\mu_z=r\exp\left[2a_\ell\int_{2M_{\rm BH}}^r\dd s\,\frac{s}{D_z(s)}\right]\,,
\end{equation}
the general solution reads
\begin{align}\label{eq:dB_generic_z}
&\delta B=\frac{1}{\mu_z}\Big[c_1\notag\\&\quad\quad\quad\quad\quad\quad-2\int_{2M_{\rm BH}}^r\dd x\,\frac{\mu_z(x)x^3}{D_z(x)}\delta V_{\rm bump}(x)\Big]\,.
\end{align}
The same near-horizon regularity condition employed in the $z=0$ case fixes $c_1=0$.

Let us consider a positive localized bump, $\delta V_{\rm bump}\geq0$, which is not identically zero. Since $\mu_z>0$ and $D_z>0$, Eq.~\eqref{eq:dB_generic_z} implies $\delta B\leq0$, with the inequality being strict once the integral receives a nonvanishing contribution.

Near the horizon, both $D_z$ and $\mu_z$ approach finite, nonvanishing values. Hence, in a sufficiently small neighborhood of the horizon, there exists a finite constant $C_h>0$ such that $\mu_zr^3/D_z\leq C_h$. Since $\delta V_{\rm bump}$ increases monotonically from zero
\begin{equation}
\int_{2M_{\rm BH}}^r\dd x\,\frac{\mu_z(x)x^3}{D_z(x)}\delta V_{\rm bump}(x)\leq C_h(r-2M_{\rm BH})\delta V_{\rm bump}(r)\,.
\end{equation}
Therefore, in the near horizon region $\delta B=o(\delta V_{\rm bump})$ and Eq.~\eqref{eq:deltarho_ani} automatically gives
\begin{equation}
\delta\rho\sim\frac{\delta V_{\rm bump}}{4\pi}\to0^{+}\,,\qquad r\to2M_{\rm BH}^{+}\,.
\end{equation}

We now consider the asymptotic region.  Defining $W_z=\mu_z/r$, one has $W_z'=2a_\ell\mu_z/D_z>0$. Moreover, Eqs.~\eqref{eq:dB_generic_z}
and~\eqref{eq:deltarho_ani} give
\begin{equation}
4\pi D_z\mu_z\delta\rho=\int_{2M_{\rm BH}}^r \dd x\,W_z(x)\frac{\dd}{\dd x}\left[x^3\delta V_{\rm bump}(x)\right]\,.
\end{equation}
The density profile must again possess a nonvanishing extremum for the same reason as in the previous paragraph, since it becomes negative at some finite radius. Consider indeed $r^3\delta V_{\rm bump}$, which vanishes at infinity if $\delta V_{\rm bump}$ decays faster than $r^{-3}$. Integrating by parts the previous equation at finite $r$ gives 
\begin{align}\label{eq:zneq0parts}
    4\pi D_z\mu_z\delta\rho&=W_z r^3 \delta V_{\rm bump}- W_zI_z\,,
\end{align}
with
\begin{equation}
    I_z={1 \over W_z}\int_{2M_{\rm BH}}^r \dd x\, W'_z(x) x^3 \delta V_{\rm bump}(x)\,.
\end{equation}
Suppose by contradiction that $\delta\rho\ge0$ everywhere, Eq.~\eqref{eq:zneq0parts} would imply
\begin{align}
0\leq I_z(r) \leq
r^3\delta V_{\rm bump}(r)\,,
\end{align}
and consequently $I(r\to\infty)=0$. 
On the other hand, 
\begin{equation}
I_z'=\frac{W_z'}{W_z^2}4 \pi D_z \mu_z \delta \rho\geq 0\,,
\end{equation}
and $I_z$ would therefore be positive and nondecreasing while approaching zero for $r\to \infty$, which is impossible. Hence $\delta\rho$ must become negative at some finite radius. 
\subsection{Regularized thin shells}
%
Another possible application of the general result of this section is to the thin-shell configuration considered in Ref.~\cite{Laeuger:2025zgb}. Thin shells are standard idealized systems in General Relativity~\cite{Israel:1966rt} and have been widely employed in studies of the equilibrium and stability of compact surface layers~\cite{Brady:1991np,Poisson:1995sv,LeMaitre:2019xez,Yang:2022gic,Pitre:2026msx}.
The construction of Ref.~\cite{Laeuger:2025zgb} consists of two Schwarzschild geometries matched across a shell located at $r=R$: the interior and exterior regions have masses $M_{\rm BH}$ and $M_{\rm BH}+\delta M$, respectively. Consequently, the mass function is discontinuous at the shell. Here, we regularize this configuration by replacing the discontinuity with a smooth transition generated by a density profile localized around $R$ and characterized by a finite width $\sigma$.
\begin{equation}
\delta \rho_{\sigma}=\frac{\delta M}{\mathcal N_\sigma}\exp\left[-\frac{(r-R)^2}{2\sigma^2}\right]\,, \label{eq:rho_gaussian_approx}
\end{equation}
where the normalization ${\mathcal N_\sigma}$ is fixed
so that
\begin{equation}
4\pi\int_{2M_{\rm BH}}^{\infty} r^2\delta\rho_\sigma(r)\dd r=\delta M\,.\end{equation}
For \(\sigma\ll R\), one has
\begin{equation}
\delta\rho_\sigma\simeq\frac{\delta M}{4\pi R^2\sqrt{2\pi}\sigma}\exp\left[-\frac{(r-R)^2}{2\sigma^2}\right]\,.
\end{equation}
Defining the corresponding perturbation of the mass function as
\begin{equation}
\delta m_\sigma=4\pi\int_{2M_{\rm BH}}^{r} \bar r^2\delta\rho_\sigma(\bar r)d\bar r \,,
\end{equation}
the thin-shell limit is recovered as
\begin{equation}
\lim_{\sigma \to 0} \delta m_\sigma
=\delta M\,\Theta(r-R)\,.
\end{equation}
A natural regular bulk analogue of an infinitesimally thin shell is obtained by imposing vanishing radial pressure,
\begin{equation}
\delta p_r = 0 \, ,
\end{equation}
since ideal thin shells carry stresses intrinsic to themselves, but no bulk pressure in the radial direction. In light of the discussion at the beginning of the section, since the regularized density profile has an extremum at $R$, here the tangential sound speed diverges, leading to a violation of causality. 

This causality violation is not cured by taking the thin-shell limit. For every finite value of $\sigma$, the Gaussian profile has a maximum at $r=R$. The sound speed should therefore be understood as a local quantity defined away from the exact maximum and then studied in the limit $r\to R$. 

Setting $x=r-R$ and using the Gaussian profile in Eq.~\eqref{eq:rho_gaussian_approx}, Eq.~\eqref{eq:cst_pr0_general} gives
\begin{equation}
c_{s,t}^2(R+x)
=
\frac{M_{\rm BH}}{2(R+x)f(R+x)}
+
\frac{M_{\rm BH}\sigma^2}
{2(R+x)^2 f(R+x)^2 x} \, .
\end{equation}
In the vicinity of the density maximum, located at $r=R$, the characteristic speed behaves as
\begin{equation}
c_{s,t}^2(R+x)
\sim
\frac{M_{\rm BH}\sigma^2}{2R^2f(R)^2}\frac{1}{x} \, ,
\qquad x\to0 \, .
\end{equation}
Thus, for every finite $\sigma$, the tangential characteristic speed has a pole at the maximum of the regularized density profile, before the distributional limit is taken. Sending $\sigma\to0$ squeezes the region containing this pathology onto the shell. Consequently, within the class of smooth local barotropic bulk realizations with vanishing radial pressure, the thin shell cannot be recovered while maintaining a finite tangential speed.

This conclusion does not depend on the Gaussian form of the regularization. Any smooth, nonnegative, localized density profile converging to a Dirac delta possesses a nonzero maximum in the localization region. For vanishing radial pressure, Eq.~\eqref{eq:cst_pr0_general} then implies a divergent tangential speed at that maximum.

\section{Conclusion}
In this work, we investigated whether localized deformations of the Regge--Wheeler potential, commonly used to study BH spectral instability, can arise from physically admissible matter distributions around a Schwarzschild BH. Our work does not question spectral instability itself, but rather constrains the interpretation of commonly used toy models in terms of realistic matter environments. Within a perturbative treatment of the matter-induced geometry, we first clarified that a secondary light ring is not generally generated by the presence of a bump in the effective potential, as long as the bump vanishes in the eikonal limit.

We then investigated whether a prescribed localized deformation can be supported by physically admissible matter. In general, environments modify the background geometry and, consequently, both the differential operator and the matter terms entering the perturbation equations. Requiring these modifications to reduce solely to an additive deformation of the effective potential imposes specific relations among the metric and stress-energy components. Thus, we show that, for physically motivated fluid models commonly considered in the literature, such as isotropic, strictly static, and irrotational fluids, as well as fluids with vanishing radial pressure, a localized bump in the effective potential either cannot be supported by a nontrivial regular matter configuration or leads to singular characteristic sound speeds. Furthermore, we show that density extrema lead to divergent characteristic speeds, an obstruction that also persists for smooth regularizations of infinitesimally thin shells. These perturbative results are consistent with the independent nonperturbative analysis of Ref.~\cite{Datta:2023zmd}, where similar pathologies were found for several matter-density profiles surrounding a BH.

Future work should test the extent to which our conclusions persist beyond the assumptions adopted here. Natural extensions include more general anisotropic fluids with nonvanishing radial pressure, $p_r\neq0$, nonbarotropic or nonlocal constitutive relations, multi-component matter, and nonperturbative configurations. Extending the analysis to the polar sector and to rotating BHs will also be important, as these cases introduce additional couplings between the gravitational and matter degrees of freedom. These directions will clarify which aspects of the obstruction found here are model-dependent and which reflect a broader limitation of pure-potential modification of the Regge--Wheeler equation.

\acknowledgments
We thank Y. Zhao for useful discussions.
This work is supported by the MUR FIS2 Advanced Grant ET-NOW (CUP:~B53C25001080001) and by the INFN TEONGRAV initiative.

\bibliography{ref}

@book{Chandrasekhar:1985kt,
    author = "Chandrasekhar, Subrahmanyan",
    title = "{The mathematical theory of black holes}",
    isbn = "978-0-19-850370-5",
    year = "1985"
}

@article{Barausse:2014tra,
    author = "Barausse, Enrico and Cardoso, Vitor and Pani, Paolo",
    title = "{Can environmental effects spoil precision gravitational-wave astrophysics?}",
    eprint = "1404.7149",
    archivePrefix = "arXiv",
    primaryClass = "gr-qc",
    doi = "10.1103/PhysRevD.89.104059",
    journal = "Phys. Rev. D",
    volume = "89",
    number = "10",
    pages = "104059",
    year = "2014"
}

@article{Ferrari:1984zz,
    author = "Ferrari, Valeria and Mashhoon, Bahram",
    title = "{New approach to the quasinormal modes of a black hole}",
    doi = "10.1103/PhysRevD.30.295",
    journal = "Phys. Rev. D",
    volume = "30",
    pages = "295--304",
    year = "1984"
}

@article{Leung:1999iq,
    author = "Leung, P. T. and Liu, Y. T. and Suen, W. M. and Tam, C. Y. and Young, K.",
    title = "{Perturbative approach to the quasinormal modes of dirty black holes}",
    eprint = "gr-qc/9903032",
    archivePrefix = "arXiv",
    doi = "10.1103/PhysRevD.59.044034",
    journal = "Phys. Rev. D",
    volume = "59",
    pages = "044034",
    year = "1999"
}

@article{Leung:1997was,
    author = "Leung, P. T. and Liu, Y. T. and Suen, W. M. and Tam, C. Y. and Young, K.",
    title = "{Quasinormal modes of dirty black holes}",
    eprint = "gr-qc/9903031",
    archivePrefix = "arXiv",
    doi = "10.1103/PhysRevLett.78.2894",
    journal = "Phys. Rev. Lett.",
    volume = "78",
    pages = "2894--2897",
    year = "1997"
}

@article{Cardoso:2019rvt,
    author = "Cardoso, Vitor and Pani, Paolo",
    title = "{Testing the nature of dark compact objects: a status report}",
    eprint = "1904.05363",
    archivePrefix = "arXiv",
    primaryClass = "gr-qc",
    doi = "10.1007/s41114-019-0020-4",
    journal = "Living Rev. Rel.",
    volume = "22",
    number = "1",
    pages = "4",
    year = "2019"
}

@article{Boyanov:2024fgc,
    author = "Boyanov, Valentin",
    title = "{On destabilising quasi-normal modes with a radially concentrated perturbation}",
    eprint = "2410.11547",
    archivePrefix = "arXiv",
    primaryClass = "gr-qc",
    doi = "10.3389/fphy.2024.1511757",
    journal = "Front. in Phys.",
    volume = "12",
    pages = "1511757",
    year = "2024"
}

@article{Maggio:2021ans,
    author = "Maggio, Elisa and Pani, Paolo and Raposo, Guilherme",
    title = "{Testing the nature of dark compact objects with gravitational waves}",
    eprint = "2105.06410",
    archivePrefix = "arXiv",
    primaryClass = "gr-qc",
    month = "5",
    year = "2021"
}

@article{Yang:2022gic,
    author = "Yang, Huan and Bonga, Beatrice and Pan, Zhen",
    title = "{Dynamical Instability of Self-Gravitating Membranes}",
    eprint = "2207.13754",
    archivePrefix = "arXiv",
    primaryClass = "gr-qc",
    doi = "10.1103/PhysRevLett.130.011402",
    journal = "Phys. Rev. Lett.",
    volume = "130",
    number = "1",
    pages = "011402",
    year = "2023"
}

@book{Morse:1946met,
  title={Methods of theoretical physics},
  author={Morse, Philip McCord and Feshbach, Herman},
  year={1946},
  publisher={Technology Press}
}

@article{DOnofrio:2026ulh,
    author = "D'Onofrio, Simone and Datta, Sayak and Maselli, Andrea",
    title = "{Axial tidal Love numbers of black holes in matter environments}",
    eprint = "2605.02633",
    archivePrefix = "arXiv",
    primaryClass = "gr-qc",
    month = "5",
    year = "2026"
}

@article{Zhao:2026eti,
    author = "Zhao, Yu-Qian and Pani, Paolo",
    title = "{Quasinormal modes and tidal responses of black holes in generic anisotropic matter environments}",
    eprint = "2606.11380",
    archivePrefix = "arXiv",
    primaryClass = "gr-qc",
    month = "6",
    year = "2026"
}

@article{Datta:2023zmd,
    author = "Datta, Sayak",
    title = "{Black holes immersed in dark matter: Energy condition and sound speed}",
    eprint = "2312.01277",
    archivePrefix = "arXiv",
    primaryClass = "gr-qc",
    doi = "10.1103/PhysRevD.109.104042",
    journal = "Phys. Rev. D",
    volume = "109",
    number = "10",
    pages = "104042",
    year = "2024"
}

@article{Maggio:2020jml,
    author = "Maggio, Elisa and Buoninfante, Luca and Mazumdar, Anupam and Pani, Paolo",
    title = "{How does a dark compact object ringdown?}",
    eprint = "2006.14628",
    archivePrefix = "arXiv",
    primaryClass = "gr-qc",
    doi = "10.1103/PhysRevD.102.064053",
    journal = "Phys. Rev. D",
    volume = "102",
    number = "6",
    pages = "064053",
    year = "2020"
}

@article{Cardoso:2016rao,
    author = "Cardoso, Vitor and Franzin, Edgardo and Pani, Paolo",
    title = "{Is the gravitational-wave ringdown a probe of the event horizon?}",
    eprint = "1602.07309",
    archivePrefix = "arXiv",
    primaryClass = "gr-qc",
    doi = "10.1103/PhysRevLett.116.171101",
    journal = "Phys. Rev. Lett.",
    volume = "116",
    number = "17",
    pages = "171101",
    year = "2016",
    note = "[Erratum: Phys.Rev.Lett. 117, 089902 (2016)]"
}

@book{Ferrari:2020nzo,
    author = "Ferrari, Valeria and Gualtieri, Leonardo and Pani, Paolo",
    title = "{General Relativity and its Applications}",
    isbn = "9781041098744; 9781041092742",
    publisher = "CRC Press",
    month = "07",
    year = "2026"
}

@article{Cardoso:2024mrw,
    author = "Cardoso, Vitor and Kastha, Shilpa and Panosso Macedo, Rodrigo",
    title = "{Physical significance of the black hole quasinormal mode spectra instability}",
    eprint = "2404.01374",
    archivePrefix = "arXiv",
    primaryClass = "gr-qc",
    doi = "10.1103/PhysRevD.110.024016",
    journal = "Phys. Rev. D",
    volume = "110",
    number = "2",
    pages = "024016",
    year = "2024"
}

@article{Regge:1957td,
    author = "Regge, Tullio and Wheeler, John A.",
    title = "{Stability of a Schwarzschild singularity}",
    doi = "10.1103/PhysRev.108.1063",
    journal = "Phys. Rev.",
    volume = "108",
    pages = "1063--1069",
    year = "1957"
}

@article{Jaramillo:2020tuu,
    author = "Jaramillo, Jos{\'e} Luis and Panosso Macedo, Rodrigo and Al Sheikh, Lamis",
    title = "{Pseudospectrum and Black Hole Quasinormal Mode Instability}",
    eprint = "2004.06434",
    archivePrefix = "arXiv",
    primaryClass = "gr-qc",
    doi = "10.1103/PhysRevX.11.031003",
    journal = "Phys. Rev. X",
    volume = "11",
    number = "3",
    pages = "031003",
    year = "2021"
}

@article{Laeuger:2025zgb,
    author = "Laeuger, Andrew and Weller, Colin and Li, Dongjun and Chen, Yanbei",
    title = "{Ringdown of a black hole surrounded by a thin shell of matter}",
    eprint = "2506.00367",
    archivePrefix = "arXiv",
    primaryClass = "gr-qc",
    doi = "10.1103/gkj4-m1c1",
    journal = "Phys. Rev. D",
    volume = "112",
    number = "8",
    pages = "084042",
    year = "2025"
}

@article{Israel:1966rt,
    author = "Israel, W.",
    title = "{Singular hypersurfaces and thin shells in general relativity}",
    doi = "10.1007/BF02710419",
    journal = "Nuovo Cim. B",
    volume = "44S10",
    pages = "1",
    year = "1966",
    note = "[Erratum: Nuovo Cim.B 48, 463 (1967)]"
}

@article{Brady:1991np,
    author = "Brady, P. R. and Louko, J. and Poisson, Eric",
    title = "{Stability of a shell around a black hole}",
    doi = "10.1103/PhysRevD.44.1891",
    journal = "Phys. Rev. D",
    volume = "44",
    pages = "1891--1894",
    year = "1991"
}

@article{Poisson:1995sv,
    author = "Poisson, Eric and Visser, Matt",
    title = "{Thin shell wormholes: Linearization stability}",
    eprint = "gr-qc/9506083",
    archivePrefix = "arXiv",
    doi = "10.1103/PhysRevD.52.7318",
    journal = "Phys. Rev. D",
    volume = "52",
    pages = "7318--7321",
    year = "1995"
}

@article{LeMaitre:2019xez,
    author = "LeMaitre, Philip and Poisson, Eric",
    title = "{Equilibrium and stability of thin spherical shells in Newtonian and relativistic gravity}",
    eprint = "1909.06253",
    archivePrefix = "arXiv",
    primaryClass = "gr-qc",
    doi = "10.1119/10.0000026",
    journal = "Am. J. Phys.",
    volume = "87",
    number = "12",
    pages = "961",
    year = "2019"
}

@article{Pitre:2026msx,
    author = "Pitre, Tristan and Schneider, Berend and Poisson, Eric",
    title = "{Self-gravitating thin shells are dynamically unstable on all angular scales}",
    eprint = "2604.05980",
    archivePrefix = "arXiv",
    primaryClass = "gr-qc",
    doi = "10.1103/ccxt-5ljs",
    journal = "Phys. Rev. D",
    volume = "113",
    number = "12",
    pages = "124055",
    year = "2026"
}

@article{Berti:2025hly,
    author = "Berti, Emanuele and others",
    editor = "Berti, Emanuele and Cardoso, Vitor and Carullo, Gregorio",
    title = "{Black hole spectroscopy: from theory to experiment}",
    eprint = "2505.23895",
    archivePrefix = "arXiv",
    primaryClass = "gr-qc",
    doi = "10.1088/1361-6382/ae59e2",
    journal = "Class. Quant. Grav.",
    volume = "43",
    number = "12",
    pages = "123001",
    year = "2026"
}

@article{ET:2025xjr,
    author = "Abac, Adrian and others",
    collaboration = "ET",
    title = "{The Science of the Einstein Telescope}",
    eprint = "2503.12263",
    archivePrefix = "arXiv",
    primaryClass = "gr-qc",
    reportNumber = "ET-0036C-25",
    month = "3",
    year = "2025"
}

@article{ET:2019dnz,
    author = "Maggiore, Michele and others",
    collaboration = "ET",
    title = "{Science Case for the Einstein Telescope}",
    eprint = "1912.02622",
    archivePrefix = "arXiv",
    primaryClass = "astro-ph.CO",
    doi = "10.1088/1475-7516/2020/03/050",
    journal = "JCAP",
    volume = "03",
    pages = "050",
    year = "2020"
}

@article{LIGOScientific:2025slb,
    author = "Abac, A. G. and others",
    collaboration = "LIGO Scientific, VIRGO, KAGRA",
    title = "{GWTC-4.0: Updating the Gravitational-Wave Transient Catalog with Observations from the First Part of the Fourth LIGO-Virgo-KAGRA Observing Run}",
    eprint = "2508.18082",
    archivePrefix = "arXiv",
    primaryClass = "gr-qc",
    reportNumber = "LIGO-P2400386",
    month = "8",
    year = "2025"
}

@article{LISA:2024hlh,
    author = "Colpi, Monica and others",
    collaboration = "LISA",
    title = "{LISA Definition Study Report}",
    eprint = "2402.07571",
    archivePrefix = "arXiv",
    primaryClass = "astro-ph.CO",
    month = "2",
    year = "2024"
}

@article{Volkel:2025jdx,
    author = {V{\"o}lkel, Sebastian H. and Dhani, Arnab},
    title = "{Quantifying systematic biases in black hole spectroscopy}",
    eprint = "2507.22122",
    archivePrefix = "arXiv",
    primaryClass = "gr-qc",
    doi = "10.1103/g6sz-dw28",
    journal = "Phys. Rev. D",
    volume = "112",
    number = "8",
    pages = "084076",
    year = "2025"
}

@article{Capuano:2025kkl,
    author = "Capuano, Lodovico and Vaglio, Massimo and Chandramouli, Rohit S. and Pitte, Chantal L. and Kuntz, Adrien and Barausse, Enrico",
    title = "{Systematic bias in LISA ringdown analysis due to waveform inaccuracy}",
    eprint = "2506.21181",
    archivePrefix = "arXiv",
    primaryClass = "gr-qc",
    doi = "10.1103/86yd-x1sl",
    journal = "Phys. Rev. D",
    volume = "112",
    number = "10",
    pages = "104031",
    year = "2025"
}

@article{Nollert:1996rf,
    author = "Nollert, Hans-Peter",
    title = "{About the significance of quasinormal modes of black holes}",
    eprint = "gr-qc/9602032",
    archivePrefix = "arXiv",
    doi = "10.1103/PhysRevD.53.4397",
    journal = "Phys. Rev. D",
    volume = "53",
    pages = "4397--4402",
    year = "1996"
}

@article{Cheung:2022rbm,
    author = "Cheung, Mark Ho-Yeuk and others",
    title = "{Nonlinear Effects in Black Hole Ringdown}",
    eprint = "2208.07374",
    archivePrefix = "arXiv",
    primaryClass = "gr-qc",
    doi = "10.1103/PhysRevLett.130.081401",
    journal = "Phys. Rev. Lett.",
    volume = "130",
    number = "8",
    pages = "081401",
    year = "2023"
}

@article{Mitman:2022qdl,
    author = "Mitman, Keefe and others",
    title = "{Nonlinearities in Black Hole Ringdowns}",
    eprint = "2208.07380",
    archivePrefix = "arXiv",
    primaryClass = "gr-qc",
    doi = "10.1103/PhysRevLett.130.081402",
    journal = "Phys. Rev. Lett.",
    volume = "130",
    number = "8",
    pages = "081402",
    year = "2023"
}

@article{Zhu:2023mzv,
    author = "Zhu, Hengrui and Ripley, Justin L. and C{\'a}rdenas-Avenda{\~n}o, Alejandro and Pretorius, Frans",
    title = "{Challenges in quasinormal mode extraction: Perspectives from numerical solutions to the Teukolsky equation}",
    eprint = "2309.13204",
    archivePrefix = "arXiv",
    primaryClass = "gr-qc",
    doi = "10.1103/PhysRevD.109.044010",
    journal = "Phys. Rev. D",
    volume = "109",
    number = "4",
    pages = "044010",
    year = "2024"
}

@article{Baibhav:2023clw,
    author = "Baibhav, Vishal and Cheung, Mark Ho-Yeuk and Berti, Emanuele and Cardoso, Vitor and Carullo, Gregorio and Cotesta, Roberto and Del Pozzo, Walter and Duque, Francisco",
    title = "{Agnostic black hole spectroscopy: Quasinormal mode content of numerical relativity waveforms and limits of validity of linear perturbation theory}",
    eprint = "2302.03050",
    archivePrefix = "arXiv",
    primaryClass = "gr-qc",
    doi = "10.1103/PhysRevD.108.104020",
    journal = "Phys. Rev. D",
    volume = "108",
    number = "10",
    pages = "104020",
    year = "2023"
}

@article{Daghigh:2020jyk,
    author = "Daghigh, Ramin G. and Green, Michael D. and Morey, Jodin C.",
    title = "{Significance of Black Hole Quasinormal Modes: A Closer Look}",
    eprint = "2002.07251",
    archivePrefix = "arXiv",
    primaryClass = "gr-qc",
    doi = "10.1103/PhysRevD.101.104009",
    journal = "Phys. Rev. D",
    volume = "101",
    number = "10",
    pages = "104009",
    year = "2020"
}

@article{Destounis:2021lum,
    author = "Destounis, Kyriakos and Macedo, Rodrigo Panosso and Berti, Emanuele and Cardoso, Vitor and Jaramillo, Jos{\'e} Luis",
    title = {{Pseudospectrum of Reissner-Nordstr{\"o}m black holes: Quasinormal mode instability and universality}},
    eprint = "2107.09673",
    archivePrefix = "arXiv",
    primaryClass = "gr-qc",
    doi = "10.1103/PhysRevD.104.084091",
    journal = "Phys. Rev. D",
    volume = "104",
    number = "8",
    pages = "084091",
    year = "2021"
}

@article{Cheung:2021bol,
    author = "Cheung, Mark Ho-Yeuk and Destounis, Kyriakos and Macedo, Rodrigo Panosso and Berti, Emanuele and Cardoso, Vitor",
    title = "{Destabilizing the Fundamental Mode of Black Holes: The Elephant and the Flea}",
    eprint = "2111.05415",
    archivePrefix = "arXiv",
    primaryClass = "gr-qc",
    doi = "10.1103/PhysRevLett.128.111103",
    journal = "Phys. Rev. Lett.",
    volume = "128",
    number = "11",
    pages = "111103",
    year = "2022"
}

@article{Courty:2023rxk,
    author = "Courty, Aubin and Destounis, Kyriakos and Pani, Paolo",
    title = "{Spectral instability of quasinormal modes and strong cosmic censorship}",
    eprint = "2307.11155",
    archivePrefix = "arXiv",
    primaryClass = "gr-qc",
    doi = "10.1103/PhysRevD.108.104027",
    journal = "Phys. Rev. D",
    volume = "108",
    number = "10",
    pages = "104027",
    year = "2023"
}

@article{Barausse:2014pra,
    author = "Barausse, Enrico and Cardoso, Vitor and Pani, Paolo",
    editor = "Ciani, Giacomo and Conklin, John W. and Mueller, Guido",
    title = "{Environmental Effects for Gravitational-wave Astrophysics}",
    eprint = "1404.7140",
    archivePrefix = "arXiv",
    primaryClass = "astro-ph.CO",
    doi = "10.1088/1742-6596/610/1/012044",
    journal = "J. Phys. Conf. Ser.",
    volume = "610",
    number = "1",
    pages = "012044",
    year = "2015"
}

@article{Berti:2022xfj,
    author = "Berti, Emanuele and Cardoso, Vitor and Cheung, Mark Ho-Yeuk and Di Filippo, Francesco and Duque, Francisco and Martens, Paul and Mukohyama, Shinji",
    title = "{Stability of the fundamental quasinormal mode in time-domain observations against small perturbations}",
    eprint = "2205.08547",
    archivePrefix = "arXiv",
    primaryClass = "gr-qc",
    doi = "10.1103/PhysRevD.106.084011",
    journal = "Phys. Rev. D",
    volume = "106",
    number = "8",
    pages = "084011",
    year = "2022"
}

@article{Yang:2024vor,
    author = "Yang, Yiqiu and Mai, Zhan-Feng and Yang, Run-Qiu and Shao, Lijing and Berti, Emanuele",
    title = "{Spectral instability of black holes: Relating the frequency domain to the time domain}",
    eprint = "2407.20131",
    archivePrefix = "arXiv",
    primaryClass = "gr-qc",
    doi = "10.1103/PhysRevD.110.084018",
    journal = "Phys. Rev. D",
    volume = "110",
    number = "8",
    pages = "084018",
    year = "2024"
}

@article{Rosato:2024arw,
    author = "Rosato, Romeo Felice and Destounis, Kyriakos and Pani, Paolo",
    title = "{Ringdown stability: Graybody factors as stable gravitational-wave observables}",
    eprint = "2406.01692",
    archivePrefix = "arXiv",
    primaryClass = "gr-qc",
    doi = "10.1103/PhysRevD.110.L121501",
    journal = "Phys. Rev. D",
    volume = "110",
    number = "12",
    pages = "L121501",
    year = "2024"
}

@article{Oshita:2024fzf,
    author = "Oshita, Naritaka and Takahashi, Kazufumi and Mukohyama, Shinji",
    title = "{Stability and instability of the black hole greybody factors and ringdowns against a small-bump correction}",
    eprint = "2406.04525",
    archivePrefix = "arXiv",
    primaryClass = "gr-qc",
    reportNumber = "YITP-24-69, IPMU24-0025, RIKEN-iTHEMS-Report-24",
    doi = "10.1103/PhysRevD.110.084070",
    journal = "Phys. Rev. D",
    volume = "110",
    number = "8",
    pages = "084070",
    year = "2024"
}

@article{Torres:2023nqg,
    author = "Torres, Theo",
    title = "{From Black Hole Spectral Instability to Stable Observables}",
    eprint = "2304.10252",
    archivePrefix = "arXiv",
    primaryClass = "gr-qc",
    doi = "10.1103/PhysRevLett.131.111401",
    journal = "Phys. Rev. Lett.",
    volume = "131",
    number = "11",
    pages = "111401",
    year = "2023"
}

\appendix
\section{Density extrema for isotropic fluids}\label{app:isotropic}
Consider an isotropic fluid, where $\delta p_t=\delta p_r=\delta p$. For such a fluid, Eq.~\eqref{eq:dpt} becomes
\begin{equation}\label{eq:dp}
    \delta p'=-\frac{M_{\rm BH}}{r^2f}\left(\delta p+\delta \rho\right) \ ,
\end{equation}
and the speed of sound reads
\begin{equation}\label{eq:speed_of_sound_iso}
    c_{s}^2=\frac{\delta p'}{\delta \rho'}=-\frac{M_{\rm BH}}{r^2f}\frac{\delta p+\delta \rho}{\delta \rho'} \ ,
\end{equation}
which diverges at $r_0$ such that $\delta \rho'(r_0)=0$ unless $\delta p(r_0)+\delta \rho(r_0)=0$. We can show that this latter condition is not satisfied if $r_0$ is the outermost nonvanishing extremum of $\delta\rho$.
Indeed, the general solution of Eq.~\eqref{eq:dp} is in the form
\begin{equation}\label{eq:dp_sol_gen}
    \delta p=f^{-1/2}\left(c_1 +\int_r^\infty \dd x \frac{M_{\rm BH} \delta \rho (x)}{x^2 f^{1/2}(x)} \right)\ .
\end{equation}
The integration constant $c_1$ can be fixed from the asymptotic behavior of the function at infinity. 
By definition $\delta p \to \delta p_0$ at infinity. On the other hand, $\delta p(r\to \infty)=0$ because the spacetime is asymptotically flat. This implies $c_1=0$, and
\begin{equation}\label{eq:dp_sol}
    \delta p=f^{-1/2}\int_r^\infty \dd x \frac{M_{\rm BH} \delta \rho (x)}{x^2 f^{1/2}(x)} \ .
\end{equation}
The above equation also determines the sign of the pressure at the outermost extremum of the density profile. Let $r_0$ be a nonvanishing extremum of $\delta\rho$, and assume that no additional extrema occur for $r>r_0$. Since $\delta\rho\to0$ at spatial infinity, the density cannot change sign in the interval $[r_0,\infty)$, otherwise it would develop an additional extremum. 
Therefore, $\delta\rho(x)$ preserves the sign of $\delta\rho(r_0)$ for all $x\geq r_0$. Since $f>0$ by definition, Eq.~\eqref{eq:dp_sol} further implies that $\delta p(r_0)$ has the same sign as $\delta\rho(r_0)$. Consequently,
\begin{equation}
\delta p(r_0)+\delta\rho(r_0)\neq0\,,
\end{equation}
and the numerator of Eq.~\eqref{eq:speed_of_sound_iso} does not vanish at $r_0$. Therefore, also in this case causality is violated whenever the density profile has an extremum. 

\end{document}